\documentclass[aps,pra,twocolumn,showpacs,groupedaddress,letterpaper]{revtex4-1}

\usepackage{graphicx}
\usepackage{subfigure}
\usepackage{color}
\usepackage{physymb}
\usepackage{enumitem}
\usepackage{mdframed}
\usepackage{framed}
\usepackage{epstopdf}

\begin{document}

\title{Upper-division student difficulties with the Dirac delta function}


\author{Bethany R. Wilcox}
\affiliation{Department of Physics, University of Colorado, 390 UCB, Boulder, CO 80309}

\author{Steven J. Pollock}
\affiliation{Department of Physics, University of Colorado, 390 UCB, Boulder, CO 80309}

\begin{abstract}
The Dirac delta function is a standard mathematical tool that appears repeatedly in the undergraduate physics curriculum in multiple topical areas including electrostatics, and quantum mechanics. While Dirac delta functions are often introduced in order to simplify a problem mathematically, students still struggle to manipulate and interpret them. To characterize student difficulties with the delta function at the upper-division level, we examined students' responses to traditional exam questions and a standardized conceptual assessment, and conducted think-aloud interviews. Our analysis was guided by an analytical framework that focuses on how students activate, construct, execute, and reflect on the Dirac delta function in the context of problem solving in physics. Here, we focus on student difficulties using the delta function to express charge distributions in the context of junior-level electrostatics. Common challenges included: invoking the delta function spontaneously, translating a description of a charge distribution into a mathematical expression using delta functions, integrating 3D or non-Cartesian delta function expressions, and recognizing that the delta function can have units.  We also briefly discuss implications of these difficulties for instruction.  
\end{abstract}

\maketitle

\section{Introduction}

Research into student difficulties at the upper-division level is a relatively new but growing area of Physics Education Research \cite{meltzer2012resource}.  Students in upper-division courses are asked to manipulate increasingly sophisticated mathematical tools as they grapple with advanced physics content.  To explore this dynamic, some of the literature around student difficulties at this level has focused on the use of mathematical tools and techniques during physics problem-solving \cite{caballero2014mathphys}.  For example, one mathematical tool that appears repeatedly throughout the undergraduate physics curriculum is the Dirac delta function (hereafter referred to as simply, the $\delta$-function). 

The $\delta$-function represents an interesting topic to investigate student difficulties for several reasons.  Despite often being discussed as a purely abstract, mathematical construct, it is rarely used in purely mathematical contexts.    In fact, physicists often introduce the $\delta$-function into a problem in order to describe or model a concrete physical system.  Additionally, it is the authors' experience that the $\delta$-function is often perceived by experts as trivial to manipulate, and is often introduced to simplify the mathematics of a problem.  However, we still observe common and persistent student difficulties using the $\delta$-function.  Moreover, while the $\delta$-function has appeared in literature around student difficulties with Fourier transforms and measurement in quantum mechanics \cite{sadaghiani2005thesis, mason2009thesis, zhu2011thesis, singh2006qcomputing}, we are unaware of any existing literature directly targeting student difficulties with the $\delta$-function.  

A physics student may be exposed to the $\delta$-function at several different points in the undergraduate curriculum.  These include its use as a tool to express volume mass or charge densities, to describe potential energies or wavefunctions in quantum mechanics, in the context of Fourier transforms, and, for more advanced situations, in the context of Green's functions.  In mathematics courses, the $\delta$-function is rigorously defined as a distribution \cite{lighthill1958fourier}, and is typically not seen by students until senior or masters level analysis courses.  Such courses are not often taken by physics majors.  Here, we will focus almost exclusively on the $\delta$-function as a tool to describe volume charge densities in the context of upper-division electrostatics as this is one of the earliest and perhaps simplest uses.  Despite the general sense of many faculty at the authors' institution that students will be familiar with the $\delta$-function by the time they reach the junior-level, we find that for some (arguably many) students junior-level electrostatics will be their first exposure to the $\delta$-function.  Given its many possible uses, we do not claim that the research presented here will span the space of all possible difficulties with the Dirac $\delta$-function; however, it will provide a sampling of the kinds of challenges that students face when manipulating $\delta$-functions. 

In this paper we utilize an analytical framework \cite{wilcox2013acer} describing the use of mathematical tools in physics problem solving to structure our investigation and analysis of student difficulties with $\delta$-functions (Sec.\ \ref{sec:methods}).  We then present our findings, including common difficulties we identified in our student population and a brief discussion of implications for instruction (Sec.\ \ref{sec:results}).  We end with limitations and future work (Sec.\ \ref{sec:discussion}).

\section{\label{sec:methods}Methods}

Problem solving at the upper-division level is often considerably longer and more complex than what is typical of the introductory level.  Making sense of students' work around these upper-division problems can be difficult because students often make multiple mistakes at different stages of a problem, which then propagate through their solutions in unpredictable ways.  To help manage this complexity, we make use of an analytical framework known as ACER (Activation, Construction, Execution, Reflection) to scaffold our analysis of student difficulties with the $\delta$-function \cite{wilcox2013acer, caballero2014mathphys}.  

\subsection{\label{sec:acer}The ACER Framework}

The ACER framework organizes the problem-solving process into four general components: \emph{activation} of mathematical tools, \emph{construction} of mathematical models, \emph{execution} of the mathematics, and \emph{reflection} on the results. These components were identified by studying expert problem solving \cite{wilcox2013acer} and are consistent with both a resources \cite{hammer2000resources} and epistemic framing \cite{bing2008thesis} perspective on the nature of knowledge.  However, the specific details of how a given mathematical tool is used in upper-division problem solving is often highly dependent on the context in which that tool is being used.  For this reason, the ACER framework was designed to be operationalized for specific mathematical tools in specific physics contexts.  The operationalization process results in a researcher-guided outline of key elements in a correct and complete solution to a particular problem or set of problems.  This process will be discussed in greater detail in Sec.\ \ref{sec:Dacer}, and additional details about the ACER framework can be found in Ref.\ \cite{wilcox2013acer}.

\subsection{\label{sec:context}Study Context}

Data for this study were largely collected from the first half of a two semester Electricity and Magnetism sequence
at the University of Colorado Boulder (CU). This course, called E\&M 1, typically covers electrostatics and magnetostatics (i.e., chapters 1-6 of Griffiths \cite{griffiths1999em}). The student population is composed of junior and senior-level physics, astrophysics, and engineering physics majors with a typical class size of 30-60 students. At CU, E\&M 1 is often taught with varying degrees of interactivity through the use of research-based teaching practices including peer instruction using clickers \cite{mazur1997pi} and tutorials \cite{chasteen2012transforming}.  

We collected data from three distinct sources for this investigation: student solutions to instructor designed questions on traditional midterm exams (N=372), responses to one question from the multiple-response Colorado Upper-division Electrostatics (CUE) Diagnostic \cite{wilcox2014cmr} (N=146 at CU, N=162 at external institutions), and two sets of think-aloud interviews (N=11).  

Exam data were collected from 7 semesters of the E\&M 1 course taught by 5 different instructors.  Of these, four were traditional research faculty, and one was a physics education researcher.  Two of these instructors, including the PER faculty member (SJP), taught the course twice during data collection.  Questions on the exams were developed solely by the instructor for that semester; however, in all cases, the question provided a mathematical expression for a charge density and asked for a description and/or sketch of the charge distribution (e.g., A1-type prompt below, see Sec.\ \ref{sec:Dacer} for the naming convention).  In one case, the students were also asked to calculate the integral of the given $\delta$-function expression.  

\begin{center}
\begin{minipage}{.75\linewidth}
{\bf \emph{A1-type Prompt:}} \vspace{-2mm}
      {\flushleft  \emph{Sketch the charge distribution:}}
      \begin{center}$\rho (x,y,z) = c \delta (x-1)$ \end{center}
      {\flushleft \emph{Describe the distribution in words too.  \\ What are the units of the constant, $c$? }}
\end{minipage}
\end{center}

The multiple-response CUE is a research-based, end-of-semester conceptual assessment.  Only one question on the CUE deals with $\delta$-functions, and it is of the same general type as the questions asked on the exams (A1-type).  However, rather than an open-ended prompt, the CUE asks a two-part, multiple-choice question.  The first part provides the student with the mathematical expression that corresponds to the \emph{mass} density of two point masses (i.e., $\rho(\vec{r})=m_1\delta(\vec{r}-\vec{r}_1)+m_2\delta(\vec{r}-\vec{r}_2)$) and asks what the integral of this expression is over all space.  The second part asks what this mass density represents physically.  This question along with the full CUE instrument can be accessed at Ref.\ \cite{q8cmr}.  CUE data from CU were collected from four semesters of the E\&M 1 course; two of these were courses for which we also have exam data.  In addition to the CU data, we also collected multiple-response CUE data from 9 courses at 7 external institutions.  These institutions range from small liberal arts colleges to large research universities.  

Think-aloud interviews were conducted in two sets performed roughly a year apart.  Interviewees were paid volunteers who had successfully completed E\&M 1 one to two semesters prior and responded to an email request for research participants.  All interviewees took the course during one of the semesters for which we have exam data.  Final course grades for interview participants ranged from D to A.  Interview protocols for both sets of interviews were designed, in part, to probe more deeply into difficulties identified in the exam solutions, and thus included one or more exam style (i.e., A1-type) questions.  In addition, the interview protocols also included questions designed to target aspects of the ACER framework not accessed by the exams and CUE.  These questions will be discussed in greater detail in the next section.  

\subsection{\label{sec:Dacer}Operationalizing ACER}

The process of operationalizing ACER is discussed in detail in Ref.\ \cite{wilcox2013acer}.  Briefly, operationalization is accomplished through a modified form of task analysis \cite{catrambone2011taps, wilcox2013acer} in which a content expert works through the problems of interest while carefully documenting their steps and mapping these steps onto the general components of the framework.  The resulting outline is then shared with other content experts and refined until consensus is reached that the key elements of the problem have been accounted for.  This expert-guided scheme then serves as a preliminary coding structure for analysis of student work and, if necessary, is further refined to accommodate unanticipated aspects of student problem-solving.  

To guide our data collection and analysis, we operationalized ACER for problems involving the use of $\delta$-functions to express the volume charge (or mass) densities of 1, 2, and 3D charge distributions.  For example, the volume charge density of a line charge running parallel to the z-axis and passing through the point (1,2,0) can be expressed as $\rho (\vec{r}) = \lambda\delta(x-1)\delta(y-2)$, where $\lambda$ is a \emph{unitful constant} representing the charge per unit length.  Expressing volume charge densities in this way is often necessary when working with the differential forms of Maxwell's Equations and can facilitate working with the integral forms of both Coulomb's Law and the Biot-Savart law.  

The elements of the operationalized ACER framework for these types of $\delta$-functions problems is described below.  Element codes are for labeling purposes only and are not meant to suggest a particular order, nor are all elements always necessary for every problem.  

\vspace{1mm}{\bf Activation of the tool:}  The first component of the framework involves identifying $\delta$-functions as the appropriate mathematical tool.  We identified two elements in the form of cues present in a prompt that are likely to activate resources associated with $\delta$-functions.  
\vspace{-2mm}
\begin{enumerate}[label={\bf A\arabic*}:, align=left] \itemsep0pt
  \item The question provides an expression for volume charge density in terms of $\delta$-functions
  \item The question asks for an expression of the \emph{volume} charge density of a charge distribution that includes point, line, or surface charges
\end{enumerate}
\vspace{-2mm}
We include element A1 because, in electrostatics, $\delta$-functions are often provided explicitly in the problem statement, effectively short-circuiting Activation.  

\vspace{1mm}{\bf Construction of the model:}  Elements in this component are involved in mapping the mathematical expression for the charge density to a verbal or pictorial representation of the charge distribution or vice versa.
\vspace{-2mm}
\begin{enumerate}[label={\bf C\arabic*}:, align=left] \itemsep0pt
  \item Relate the shape of the charge distribution to the coordinate system and number of $\delta$-functions
  \item Relate the location of the charges with the argument(s) of the $\delta$-function(s)
  \item Establish the need for and/or physical meaning of the \emph{unitful constant} in front of the $\delta$-function
\end{enumerate}
\vspace{-2mm}
For problems that also require integration of the $\delta$-function (e.g., to find total charge from $\rho(\vec{r})$) there are additional elements in construction related to setting up this integral.    
\vspace{-2mm}
\begin{enumerate}[label={\bf C\arabic*}:, align=left] \itemsep0pt
\setcounter{enumi}{3}
  \item Express a differential volume element consistent with the geometry of the charge distribution
  \item Select limits of integration consistent with the differential volume element and region of interest
\end{enumerate}
\vspace{-2mm}
Alternatively, for sufficiently simple charge distributions, one can bypass setting up and computing integrals (elements C4 \& C5) by using the physical meaning of the unitful constant to simply state the total charge (e.g., Q(uniform, spherical shell) = $\sigma*4\pi R^2$).  

\vspace{1mm}{\bf Execution of the mathematics:}  This component of the framework deals with elements involved in executing the mathematical operations related to the $\delta$-function.  Because this component deals with actually performing mathematical operations, these elements are specific to problems requiring integration of the $\delta$-function.  
\vspace{-2mm}
\begin{enumerate}[label={\bf E\arabic*}:, align=left] \itemsep0pt
  \item Execute (multivariable) integrals that include one or more (potentially multidimensional) $\delta$-functions
\end{enumerate}
\vspace{-2mm}
When the results of the integrals in E1 must be simplified for interpretation, Execution would include a second element relating to algebraic manipulation; however, none of the integrals included in this study elicited or required significant algebraic manipulation.

\vspace{1mm}{\bf Reflection on the result:} The final component includes elements related to checking and interpreting aspects of the solution, including intermediate steps and the final result.  While many different techniques can be used to reflect on a physics problem, the following three are particularly common when dealing with $\delta$-functions.  
\vspace{-2mm}
\begin{enumerate}[label={\bf R\arabic*:}, align=left] \itemsep0pt
  \item Check/determine the units of all relevant quantities (e.g., Q, $\rho$, the unitful constant)
  \item Check that the physical meaning of the unitful constant is consistent with its units and the units of all other quantities
  \item Verify that the value of the charge in a region is consistent with expectations
\end{enumerate}
\vspace{-2mm}
While the first two elements are similar, we consider element R2 to be a distinct and potentially more sophisticated reflection task in that it is seeking consistency between the student's physical interpretation of the unitful constant and its units.  The distinction between elements R1 and R2 was motivated in part by preliminary analysis of student work, which demonstrated that the link between the units of the constant and its physical interpretation may not be explicit for many students (see Sec.\ \ref{sec:reflection}).

The operationalization of the ACER framework for $\delta$-functions revealed that the standard questions typically asked on midterm exams (A1-type) do not capture all aspects of problem-solving with the $\delta$-function.  In particular, this type of question bypasses Activation at anything more than the most surface level assessment of whether a student recognize the $\delta$-function.  To probe Activation more deeply, we began the first set of think-aloud interviews with a question that provided a description of the charge distribution and asked for a mathematical expression for the charge density (e.g., A2-type prompt below).  This A2-type interview prompt also provides a different perspective on Construction than the A1-type exam prompt by requiring students to generate a multidimensional $\delta$-function expression rather than just interpreting it.  However, participants in the first interview set often failed to activate the $\delta$-function in response to the A2-type prompt (see Sec.\ \ref{sec:results}) and thus never moved on to deal with Construction.  In order to target the Construction component more clearly, the second set of interviews directly prompted students to use $\delta$-functions to express the charge density.    

\begin{center}
    \begin{minipage}{.75\linewidth}
     {\bf \emph{A2-type Prompt:}}\vspace{-2mm}
      \flushleft  \emph{What is the volume charge density of an infinitely long, linear charge distribution running parallel to the z-axis and passing through the point (1,2,0)?  Be sure to define any new symbols you introduce.   \\ }
      \emph{Sketch this charge distribution. }
    \end{minipage} 
\end{center}

Additionally, only one of the seven exam questions asked students to integrate an expression containing a $\delta$-function.  To further probe students ability to integrate the $\delta$-function (element E1), both interview protocols included questions that prompted students to calculate the total charge within a finite region of space.  The second set of interviews also ended by asking students to perform a set of context-free integrations of various $\delta$-function expressions (below) in order to more clearly investigate the Execution component.  These particular integral expressions were each designed to target a specific difficulty we anticipated students might have with the procedural aspects of integrating the $\delta$-function.  While all integrals were presented without a physical context, in the case of integral \emph{d)} students were also asked if they could come up with a physical situation in which they might set up this integral.

\begin{center}
\begin{minipage}{.75\linewidth}
    {\bf \emph{Context-free Integral problems:}}\vspace{-2mm}
    \flushleft 
    \begin{enumerate}[label=\emph{\alph*)},align=left] \itemsep0pt
      \item $\int\limits_\infty^{-\infty} \delta(x)dx$ 
      \item $\int\limits_\infty^{-\infty} x\delta(x)dx$ 
      \item $\int\limits_0^{10} [a \delta(x-1)+b\delta(x+2)]dx$ 
      \item $\iiint a \delta(r-r')r^2 sin(\theta)dr d\phi d\theta$ 
    \end{enumerate}
\end{minipage}
\end{center}

\section{\label{sec:results}Findings}

This section presents the identification and analysis of common student difficulties with the Dirac $\delta$-function organized by component and element of the operationalized ACER framework (Sec.\ \ref{sec:Dacer}).  

\subsection{\label{sec:activation}Activation of the tool}

Elements A1 and A2 of the framework represent two different types of of prompts that can cue students' to activate resources related to the $\delta$-function.  In the case of A2-type prompts, the student must first recognize that the $\delta$-function is the appropriate mathematical tool before they can correctly answer the question.  However, for A1-type questions, the $\delta$-function is given as part of the prompt, effectively short-circuiting Activation and providing little information about students ability to recognize \emph{when} the $\delta$-function is appropriate.  

None of the instructor-written exams included A2-type questions.  Instead, this element was specifically targeted during the first of the two interview sets.  When presented with the A2-type prompt shown in Sec.\ \ref{sec:Dacer}, 2 of 5 interviewees spontaneously suggested using $\delta$-functions.  The remaining three participants all expressed confusion at being asked to provide a volume charge density for a line of charge.  Two of these students attempted to reconcile this by defining an arbitrary cylindrical volume, $V$, around the line charge and using $\rho=Q/V$.  This strategy, while incorrect, represents a reasonable attempt to make sense of the problem in lieu of the $\delta$-function.  Later in the interview, all five of these students were presented with the $\delta$-functions expression for a linear charge density and all but one correctly interpreted this expression as describing a line charge.  This result suggests that even immediately after completing a junior-level electrostatics course, many students may have difficulty recognizing \emph{when} the $\delta$-function is the appropriate mathematical tool even when they are able to provide a correct physical interpretation of it.  

Three participants in the second interview set had just completed one semester of upper-division quantum mechanics at the time of the interview.   To investigate the context-dependent nature of student Activation of the $\delta$-function, we began these students' interviews by asking for a mathematical expression for the potential of a finite square well in the limit that the well became \emph{very} narrow and \emph{very} deep.  Two of these three participants spontaneously suggested the $\delta$-function as the appropriate tool.  Both of these students explicitly focused on the use of the words ``\emph{very} narrow and \emph{very} deep'' just before suggesting the $\delta$-function.  The third student initially attempted to write the potential as a piecewise function but brought up the $\delta$-function when explicitly told there was a more compact way to represent the potential without using a piecewise notation.  Successful activation of the $\delta$-function in the quantum case seemed to be linked to the high degree of similarity between the visual representation of a deep, narrow potential well and the commonly used graphical representation of $\delta(x)$.  

\subsection{\label{sec:construction}Construction of the model}

The Construction component deals with mapping between the physics and mathematics of a problem.  In the case of $\delta$-functions, this mapping can be done in two directions: from mathematics to physical description, or from physical description to mathematics.  The A1-type prompts (see Sec.\ \ref{sec:context}) used on exams, the CUE, and in interviews investigated students' ability to translate a mathematical expression for the charge density into a physical description of the charge distribution.  As part of this process, students needed to connect the coordinate system and number of $\delta$-functions in the given expression to the shape of the charge distribution (element C1).  For example, the expression $\rho(x,y,z) = c\delta(x-1)$ contains one Cartesian $\delta$-function and thus represents an infinite plane of charge.  Roughly a quarter of students (23\%, N=87 of 372) had an incorrect shape on exams.  The most common error was misidentifying volume charge densities that included 1 or 2 $\delta$-functions as point charges (62\%, N=54 of 87).  

On the CUE diagnostic administered at the end of the semester, the fraction of students selecting the incorrect shape was roughly a third (35\%, N=51 of 146).  This trend of selecting the incorrect shape is slightly more pronounced in student populations beyond CU who have taken the multiple-response CUE (49\%, N=79 of 162, 7 institutions) The most common error for this question was misidentifying the given point mass density ($\rho(\vec{r}) = m_1 \delta^3(\vec{r} - \vec{r}_1)$ ) as a solid sphere or spherical shell with radius $\vec{r}_1$ (76\%, N=99 of 130 incorrect responses, all institutions).  This finding is interesting because, given students' tendency to misidentify charge densities as point charges on exams, we might expect that they would be more successful at identifying point charge densities.  However, we hypothesize that the prevalence of this particular error may have been exacerbated by the apparent similarity between the expressions for this point mass density and that of a spherical shell (i.e., $\rho(\vec{r})=m_1\delta(r-r_1)$), which are distinguished only by the presence of the cube on the $\delta$-function and vectors in the argument.  

\begin{figure}
\includegraphics[width=3.4in, height=2.5in]{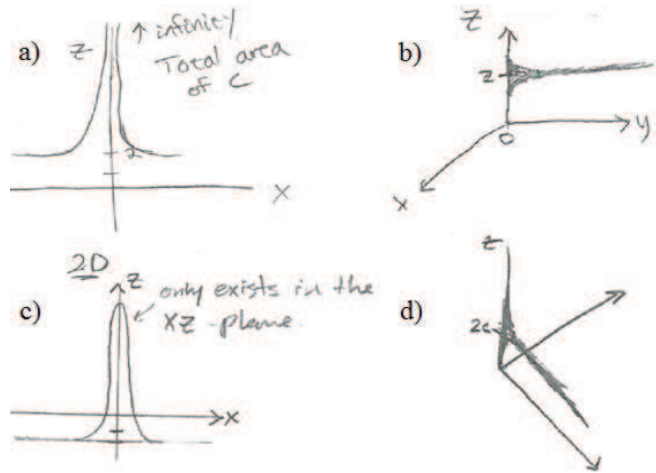}
\caption{Examples of the `spike' representation used by 5\% of exam students and 2 of 11 interview students.  In this case, students were prompted to sketch the charge distribution described by $\rho(x,y,z)=c\delta(x)\delta(z-2)$. }\label{fig:spike}
\end{figure}

In interviews, 9 of 11 participants correctly identified the shape of one or more charge distributions from the mathematical expressions for the charge density.  The remaining 2 students both sketched the charge distribution on 3D Cartesian axes as a very narrow spike originating at a point consistent with the argument of the $\delta$-function and extending upwards parallel to one of the axes (e.g., see Fig.\ \ref{fig:spike}).  This `spike' representation was also seen in a small number of the exam solutions (5\%, N=17 of 311).  The spike drawn by these students is highly reminiscent of the 1D graphical representation of $\delta(x)$ as an infinitely tall and thin Gaussian distribution at $x=0$ that is commonly used when first defining the $\delta$-function.  Students who draw this `spike' when sketching the charge distribution may be attempting to apply this 1D representation of the $\delta$-function to a 3D sketch.  The examples shown in Fig.\ \ref{fig:spike} have a number of additional interesting features.  For example, the student in Fig.\ \ref{fig:spike}(d) places the spike as $z=2c$ rather than $z=2$.  Additionally, the spikes are not always parallel to the same axis despite all having the same prompt.  These features may warrant further investigation in future studies; however, as they were observed in only a small number of exam solutions and in none of the interviews, we will not discuss them in further detail here.  

The interviews also included an A2-type prompt (see Sec.\ \ref{sec:Dacer}) to explore students' ability to translate a physical description of the charge distribution into a mathematical expression for the charge density.  This process requires students to use the geometry of the charge distribution to select an appropriate coordinate system and number of $\delta$-functions.  Of the eight interview participants who attempted to use $\delta$-functions in response to the A2-type prompt, three were able to correctly express the line charge density as the product of two 1D Cartesian $\delta$-functions.  Four of the remaining five students instead attempted to use a single $\delta$-function whose argument was the difference between two vectors, for example $\rho \propto \delta(\vec{r}-\vec{r}_1)$ where $\vec{r}_1 = (1,2,z)$.  These students did not explicitly acknowledge this as a 3D $\delta$-function either verbally or by cubing the $\delta$-function.  Three of these students also described the line charge as a continuous sum of point charges and integrated their $\delta$-function expression over all $z$.  This strategy reflects a fundamentally reasonable physical interpretation of the situation that can be used to construct a correct expression for the line charge density; however, none of these students was able to fully leverage this interpretation to correctly express the charge density.  

Determining the location of the charge distribution (element C2) was not a significant stumbling block for students in this study. None of the interview students and a tenth of the exam students (10\%, N=39 of 372) drew an incorrect position for the distribution. The most common errors were switching the signs of the coordinates (36\%, N=14 of 39, e.g., locating the plane described by $\rho(x,y,z)=c\delta(x-1)$ at x=-1) or having the wrong orientation of line or plane distributions (38\%, N=15 of 39). However, all relevant exam questions in this study dealt with locating charge distributions described by $\delta$-functions in Cartesian coordinates, and it is possible that student difficulties with element C2 would be more significant for other geometries or more abstract notation.

The third element in Construction relates to the need for a unitful constant in the expression for $\rho(\vec{r})$.  For A1-type questions, this constant is provided as part of the given expression for the charge density, and fully interpreting this expression requires recognizing the physical significance of this constant.  For example, in the expression $\rho(x,y,z)=c\delta(x-1)$, the constant, $c$, represents the charge per unit area on the surface of the plane.  Roughly a quarter of the exam students (27\%, N=70 of 255) presented with an arbitrary constant spontaneously commented on its physical meaning and most of these (81\%, N=57 of 70) had a correct interpretation. This fraction should be interpreted as a lower bound as additional students may have recognized the constant's physical significance but did not explicitly write it down on the exam. The interviews suggest that a students' interpretation of the constant can be facilitated or impeded by their identification of its units. This dynamic will be discussed in greater detail in relation to the Reflection component (Sec.\ \ref{sec:reflection}).

The A2-type prompts used in interviews, on the other hand, do not include the unitful constant but instead require that a student recognize the need for this constant independently.  Two of eight interviewees did not spontaneously include a multiplicative constant in their expression for the charge density.  Alternatively, three of the remaining participants recognized the need for a multiplicative factor but expressed this factor as $\rho(\vec{r})$ (e.g., \emph{volume charge density} $= \rho(\vec{r})\delta(x-1)\delta(y-2)$).  These students appeared to be interpreting the $\rho(\vec{r})$ term as representing only the magnitude of the charge density rather than the magnitude at a specific point in space.  This also suggests that these students are not treating $\rho$ as the quantity defined by convention to be volume charge density and are instead treating it as the quantity conventionally defined as $\lambda$ (i.e., line charge density).  Ultimately, four of the eight participants included a multiplicative constant \emph{and} articulated a correct physical interpretation of that constant before moving on from this question.  

The final two elements in Construction (elements 4 \& 5) are not specific to questions involving $\delta$-functions, but rather apply to any physics problem involving multivariable integration.  These elements deal with expressing the differential volume element and selecting limit of integration.  We have previously examined student work around these two elements in the context of Coulomb's law integrals \cite{wilcox2013acer} and observed a number of difficulties.  However, only a small number of students (21\%, N=12 of 56) struggled with either the Cartesian differential volume element or limits of integration in the single semester where the exam question asked for total charge in a finite region of space.  Furthermore, only one of eleven interviewees had clear difficulties expressing the differential charge element in spherical coordinates, and one other selected incorrect limits for their Cartesian integral.  Thus, setting up integrals for the relatively simple Cartesian and spherical geometries used in this study was not a significant challenge for our upper-division students. 

\subsection{\label{sec:execution}Execution of the mathematics}

The Execution component of the framework deals with the procedural aspects of working with mathematical tools in physics.  The exams provide limited insight into this component as students were asked to actually calculate an integral of a $\delta$-function in only one semester.  In this case, the students were given the expression for 3 point charges and asked to calculate $\int \rho(\vec{r})d\tau$.  Roughly a quarter of the students (27\%, N=15 of 56) made significant mathematical errors related to the $\delta$-function while executing this integral (element E1). The most common error (73\%, N=11 of 15) amounted to a variation of equating the integral of the $\delta$-function with the integral of the zero point of its vector argument (e.g., $ \int \delta^3(\vec{r}-\vec{r}_1) d\tau = \int \vec{r}_1 d\tau$).  Similarly, more than a third of students (36\%, N=53 of 146 at CU, 41\%, N=66 of 162 at external institutions) selected an incorrect value for the integral of a point mass density in response to the CUE question.  The most common error (81\%, N=43 of 53 at CU, 68\% N=45 of 66 at external institutions) was equating the integral of the $\delta$-function with the value (vector or magnitude) of $\vec{r}$ at which the argument was zero (e.g., $\int \delta^3(\vec{r}-\vec{r}_1) d\tau = \vec{r}_1 $).  This issue is different from (though potentially related to) the most common issue seen in the exam; however, none of the response options on the multiple-choice CUE question matched the outcome of the incorrect setup seen commonly on exams.  Ultimately, these types of errors indicate that students have correctly recognized that the $\delta$-function somehow picks out the value $\vec{r}=\vec{r}_1$ but apply this reasoning in an ad hoc fashion in the integration process.  

The first set of interviews investigated Execution in the context of calculating the total charge on a uniformly charged spherical shell.  Two of the five participants were not able to complete this calculation and explicitly stated this was because they could not recall how to integrate the $\delta$-function.  When asked for the value of $\int_{-\infty}^{\infty} \delta(x)dx$ these participants guessed it would be $x$ or $1/x$.  Extending this logic into three dimensions, this response appears consistent with some of the common incorrect responses on the CUE and exam questions (e.g., $\int \delta^3(\vec{r}-\vec{r}_1) d\tau = \vec{r}_1 $ or $1/|\vec{r}_1|$).  

The second interview set (N=6) targeted the first element in Execution differently by asking students to perform the context-free integrations shown in Sec.\ \ref{sec:Dacer}. Two students stated that the integral in b) would be equal to x without evaluating this expression at x = 0, but none of the six participants had difficulty with the integrals in a) or c). Moreover, the 3D $\delta$-functions used on the exam and CUE seemed to evoke different, and potentially more fundamental, difficulties than the 1D $\delta$-functions used in interviews. Three of six interviewees also evaluated the $r$ integral in part d) in the following way, $\int \delta(r-r')r^2 dr=\frac{1}{3}r'^3$, despite correctly executing parts a)-c). Two of these students explicitly stated that the effect of the $\delta$-function was only to pick out the value $r=r'$.  This result is consistent with that from the exams and CUE, and suggests that a significant fraction of our upper-division students have internalized the idea that the $\delta$-function picks out a particular value of the variable, but do not recognize the other impacts of the $\delta$-functions on the result of the integral, particularly when dealing with 3D or non-Cartesian $\delta$-functions.  

\subsection{\label{sec:reflection}Reflection on the results}

The Reflection component deals with aspects of the problem-solving process related to interpreting and checking intermediate steps and the final solution.  For the questions used in this study, one of the most powerful tools available for checking and interpreting the various $\delta$-function expressions is looking at units (elements R1 and R2).  In particular, for A1-type problems, looking at the units of the given constant (e.g., the $c$ in $\rho(x,y,z)=c\delta(x-1)$) can facilitate interpretation of that constant's physical meaning.  Five of the seven exam prompts explicitly asked students to comment on the units of the given constant and two-thirds of the students (70\%, N=178 of 255) responded with the correct units (element R1).  

Beyond just commenting on the units of the given constant, we would like our students to recognize the physical interpretation of this constant (element C3), and ensure that the units and physical interpretation are consistent (element R2).  However, it was often difficult to assess if students did this on the exams, in part because only a quarter of our students (N=70 of 255) explicitly commented on the physical interpretation of the unitful constant (element C3, see Sec.\ \ref{sec:construction}).  Most of these students (83\%, N=58 of 70) also provided units for the constant that were consistent with their physical interpretation of its meaning.  However, a third of the students (32\%, N=82 of 255) gave units that were inconsistent with the shape they identified regardless of whether they commented on the constant's physical meaning.  Some of these students had the correct shape but incorrect units (40\%, N=33 of 82), and others had the incorrect shape but correct units (30\%, N=25 of 82).  For an expert, the units of the constant, its physical meaning, and the shape of the charge distribution are tightly linked; however, these results suggest that this relationship may still be developing for the students.

Interviews offer additional insight into the connection between the units and physical interpretation of the unitful constant.  When prompted to comment on units, 9 of 11 participants explicitly argued (incorrectly) that the $\delta$-function is unitless and thus, regardless of the geometry of the charge distribution, the units of the constant must be $C/m^3$. This argument was often justified by the statement that the $\delta$-function was `just a mathematical thing' and thus did not have units.  The belief that the $\delta$-function is unitless was surprisingly persistent, and in some cases interfered with students' physical interpretation of the unitful constant.  For example, four of these nine students had articulated a correct physical argument for the units of the constant prior to arguing that the $\delta$-function was unitless.  Two of these students abandoned their previous physical argument in favor of a unitless $\delta$-function, and the other two were unable to reconcile their intuition about the constant's units with their belief that the $\delta$-function should be unitless.  Ultimately, only 2 of the 9 students were able to independently convince themselves that the $\delta$-function had units of $1/m$ (in this case).  The other 7 students were only convinced when specifically prompted by the interviewer to consider the units of $dx$ in the expression $\int\delta(x)dx=1$.   

The third element in Reflection deals with using the expression for the charge density to calculate the total charge within a finite region in order to ensure this value is consistent with expectations.  While this is something that we saw experts do spontaneously, almost no students (N=3 of 255) explicitly executed such a check on the exams when not prompted to calculate total charge.  Similarly, only one of the interview students spontaneously attempted to calculate total charge in a region to gain confidence in his answer.  Interview students were also explicitly prompted to calculate total charge and to consider the units of this expression (element R1) later in the interview.  This process helped 4 of the 9 students who argued that the $\delta$-function was unitless to realize there was a problem with the units of their expression.  These findings underscore the potential value of these reflective practices (element R1-3) both in terms of catching errors and facilitating interpretation, but also suggest that our students rarely engaged in reflection spontaneously.  

\subsection{\label{sec:implications}Implications for Instruction}

This study was not designed to investigate the impacts of different instructional strategies or curricular materials on the prevalence or persistence of students' difficulties with the Dirac $\delta$-function; however, our findings do suggest several implications for teaching or using the $\delta$-function at the upper-division level.  For example, instructors should be aware that students are unlikely to encounter the $\delta$-function in their required mathematics courses, and that it may or may not be covered in a math methods course run within a Physics department.  Thus, the assumption that all students in a junior-level course will be aware of the $\delta$-function and its properties may not be justified.  Students' struggles with the procedural aspects of integrating more complex $\delta$-function expressions may be one manifestation of this lack of sufficient prior experience with the $\delta$-function.  These students may benefit from additional opportunities to practice integrating both 1D and 3D delta functions and in multiple coordinate systems.  

Additionally, canonical $\delta$-functions questions rarely, if ever, require a student to consider \emph{when} the $\delta$-function is an appropriate tool.  Questions that describe a charge distribution and ask for an expression for the charge density can provide a baseline assessment of student ability to activate the $\delta$-function when not prompted explicitly.  This type of question also addresses another finding, that constructing a mathematical expression for the charge density is a distinct and potentially more challenging task for our students than interpreting that same expression.  As the former task is arguably the more authentic, students may benefit from additional opportunities to construct various $\delta$-function expressions in multiple coordinate systems.  

The belief that the $\delta$-function is unitless was a surprisingly prevalent and persistent idea.  This belief may be exacerbated by presenting the $\delta$-function as a purely abstract mathematical construct.  Moreover, the idea of a unitless $\delta$-function can interfere with students' interpretation of the unitful constant.  To facilitate student reflection on problems involving the $\delta$-function, specific emphasis should be placed not only on the fact that the $\delta$-function \emph{can} have units, but also on \emph{how} to determine them based on the argument of the $\delta$-function.  

\section{\label{sec:discussion}Conclusions}

This paper contributes to the limited body of research on student difficulties with the Dirac $\delta$-function by presenting an application of the ACER framework to guide analysis of student problem-solving with the $\delta$-function in the context of expressing charge densities mathematically in junior-level electrostatics.  The ACER framework provided an organizing structure for our analysis that helped us identify nodes in students' work where key difficulties appeared. It also informed the development of interview protocols that targeted aspects of student problem solving not accessed by traditional exams, particularly around activating the $\delta$-function as the appropriate tool and executing integrals of various $\delta$-function expressions.  

Our upper-division students encountered a number of issues when using/interpreting the $\delta$-function. These difficulties represent a subset of students' difficulties with the $\delta$-function and may not include issues that might arise from its uses in contexts outside of electrostatics.  For example, our students often struggled to activate the $\delta$-functions as the appropriate mathematical tool when not explicitly prompted.  As our upper-division students progress forward through the undergraduate and graduate curriculum, it becomes increasingly important that they be able to recognize situations in which particular tools will be useful.  Additionally, we found that students have a greater degree of difficulty translating a verbal description of a charge distribution into a mathematical formula for volume charge density than the reverse process.  While interpreting a mathematical formula for the charge density is a valuable skill for our physics majors, the ability to construct that same formula from scratch is potentially an even more valuable skill that our physics majors are likely to use in the future.  

Our students also encountered difficulties with the procedural aspects of integrating 3D and/or non-Cartesian $\delta$-functions despite often recognizing that the $\delta$-function picks out the value of the integral at a single point.  These difficulties manifest both in solving integrals embedded in a physics context and those presented in a purely mathematical context.  Finally, our students demonstrated significant difficulty determining the units of the $\delta$-function, thus limiting their ability to interpret or check their expressions for the charge density.  

Additional work is needed to identify student difficulties when utilizing the $\delta$-function in other contexts such as quantum mechanics, Fourier transforms, and Green's functions.  Such investigations could also provide a longitudinal perspective on the growth of student understanding over time, allowing researchers and instructors to focus their efforts on addressing those difficulties that are most common and most persistent throughout the physics curriculum.

\begin{acknowledgments}
Particular thanks to the PER@C group and Marcos D. Caballero for their feedback.
This work was funded by NSF-CCLI Grant DUE-1023028 and an NSF Graduate Research Fellowship under Grant No. DGE 1144083.
\end{acknowledgments}

\bibliography{master-refs-10-18-14}

\begin{thebibliography}{16}
\expandafter\ifx\csname natexlab\endcsname\relax\def\natexlab#1{#1}\fi
\expandafter\ifx\csname bibnamefont\endcsname\relax
  \def\bibnamefont#1{#1}\fi
\expandafter\ifx\csname bibfnamefont\endcsname\relax
  \def\bibfnamefont#1{#1}\fi
\expandafter\ifx\csname citenamefont\endcsname\relax
  \def\citenamefont#1{#1}\fi
\expandafter\ifx\csname url\endcsname\relax
  \def\url#1{\texttt{#1}}\fi
\expandafter\ifx\csname urlprefix\endcsname\relax\def\urlprefix{URL }\fi
\providecommand{\bibinfo}[2]{#2}
\providecommand{\eprint}[2][]{\url{#2}}

\bibitem[{\citenamefont{Meltzer and Thornton}(2012)}]{meltzer2012resource}
\bibinfo{author}{\bibfnamefont{D.~E.} \bibnamefont{Meltzer}} \bibnamefont{and}
  \bibinfo{author}{\bibfnamefont{R.~K.} \bibnamefont{Thornton}},
  \emph{\bibinfo{title}{Resource letter alip--1: Active-learning instruction in
  physics}}, \bibinfo{journal}{Am. J. Phys.} \textbf{\bibinfo{volume}{80}},
  \bibinfo{pages}{478} (\bibinfo{year}{2012}).

\bibitem[{\citenamefont{{Caballero} et~al.}(2014)\citenamefont{{Caballero},
  {Wilcox}, {Doughty}, and {Pollock}}}]{caballero2014mathphys}
\bibinfo{author}{\bibfnamefont{M.~D.} \bibnamefont{{Caballero}}},
  \bibinfo{author}{\bibfnamefont{B.~R.} \bibnamefont{{Wilcox}}},
  \bibinfo{author}{\bibfnamefont{L.}~\bibnamefont{{Doughty}}},
  \bibnamefont{and} \bibinfo{author}{\bibfnamefont{S.~J.}
  \bibnamefont{{Pollock}}}, \emph{\bibinfo{title}{{Unpacking Students' Use of
  Mathematics in Upper-division Physics}}}, \bibinfo{journal}{ArXiv e-prints}
  (\bibinfo{year}{2014}), \eprint{1409.7660}.

\bibitem[{\citenamefont{Sadaghiani}(2005)}]{sadaghiani2005thesis}
\bibinfo{author}{\bibfnamefont{H.}~\bibnamefont{Sadaghiani}},
  \bibinfo{type}{Ph.d.}, \bibinfo{school}{The Ohio State University}
  (\bibinfo{year}{2005}).

\bibitem[{\citenamefont{Mason}(2009)}]{mason2009thesis}
\bibinfo{author}{\bibfnamefont{A.}~\bibnamefont{Mason}}, \bibinfo{type}{Ph.d.
  dissertation}, \bibinfo{school}{University of Pittsburgh}
  (\bibinfo{year}{2009}).

\bibitem[{\citenamefont{Zhu}(2011)}]{zhu2011thesis}
\bibinfo{author}{\bibfnamefont{G.}~\bibnamefont{Zhu}}, \bibinfo{type}{Ph.d.},
  \bibinfo{school}{University of Pittsburgh} (\bibinfo{year}{2011}).

\bibitem[{\citenamefont{Singh}(2006)}]{singh2006qcomputing}
\bibinfo{author}{\bibfnamefont{C.}~\bibnamefont{Singh}}, in
  \emph{\bibinfo{booktitle}{Physics Education Research Conference 2006}}
  (\bibinfo{address}{Syracuse, New York}, \bibinfo{year}{2006}), vol.
  \bibinfo{volume}{883} of \emph{\bibinfo{series}{PER Conference Invited
  Paper}}, pp. \bibinfo{pages}{42--45}.

\bibitem[{\citenamefont{Lighthill}(1958)}]{lighthill1958fourier}
\bibinfo{author}{\bibfnamefont{M.~J.} \bibnamefont{Lighthill}},
  \emph{\bibinfo{title}{An introduction to Fourier analysis and generalised
  functions}} (\bibinfo{publisher}{Cambridge University Press},
  \bibinfo{year}{1958}).

\bibitem[{\citenamefont{Wilcox et~al.}(2013)\citenamefont{Wilcox, Caballero,
  Rehn, and Pollock}}]{wilcox2013acer}
\bibinfo{author}{\bibfnamefont{B.~R.} \bibnamefont{Wilcox}},
  \bibinfo{author}{\bibfnamefont{M.~D.} \bibnamefont{Caballero}},
  \bibinfo{author}{\bibfnamefont{D.~A.} \bibnamefont{Rehn}}, \bibnamefont{and}
  \bibinfo{author}{\bibfnamefont{S.~J.} \bibnamefont{Pollock}},
  \emph{\bibinfo{title}{Analytic framework for students’ use of mathematics
  in upper-division physics}}, \bibinfo{journal}{Phys. Rev. ST Phys. Educ.
  Res.} \textbf{\bibinfo{volume}{9}}, \bibinfo{pages}{020119}
  (\bibinfo{year}{2013}),
  \urlprefix\url{http://link.aps.org/doi/10.1103/PhysRevSTPER.9.020119}.

\bibitem[{\citenamefont{Hammer}(2000)}]{hammer2000resources}
\bibinfo{author}{\bibfnamefont{D.}~\bibnamefont{Hammer}},
  \emph{\bibinfo{title}{Student resources for learning introductory physics}},
  \bibinfo{journal}{Am. J. Phys.} \textbf{\bibinfo{volume}{68}},
  \bibinfo{pages}{S52} (\bibinfo{year}{2000}).

\bibitem[{\citenamefont{Bing}(2008)}]{bing2008thesis}
\bibinfo{author}{\bibfnamefont{T.~J.} \bibnamefont{Bing}},
  \bibinfo{type}{Dissertation}, \bibinfo{school}{University of Maryland}
  (\bibinfo{year}{2008}).

\bibitem[{\citenamefont{Griffiths}(1999)}]{griffiths1999em}
\bibinfo{author}{\bibfnamefont{D.~J.} \bibnamefont{Griffiths}},
  \emph{\bibinfo{title}{Introduction to electrodynamics}}
  (\bibinfo{publisher}{Prentice Hall}, \bibinfo{year}{1999}), ISBN
  \bibinfo{isbn}{9780138053260},
  \urlprefix\url{http://books.google.com/books?id=M8XvAAAAMAAJ}.

\bibitem[{\citenamefont{Mazur}(1997)}]{mazur1997pi}
\bibinfo{author}{\bibfnamefont{E.}~\bibnamefont{Mazur}},
  \emph{\bibinfo{title}{Peer Instruction: A User's Manual}}, Series in
  Educational Innovation (\bibinfo{publisher}{Prentice Hall},
  \bibinfo{address}{Upper Saddle River}, \bibinfo{year}{1997}).

\bibitem[{\citenamefont{Chasteen et~al.}(2012)\citenamefont{Chasteen, Pollock,
  Pepper, and Perkins}}]{chasteen2012transforming}
\bibinfo{author}{\bibfnamefont{S.~V.} \bibnamefont{Chasteen}},
  \bibinfo{author}{\bibfnamefont{S.~J.} \bibnamefont{Pollock}},
  \bibinfo{author}{\bibfnamefont{R.~E.} \bibnamefont{Pepper}},
  \bibnamefont{and} \bibinfo{author}{\bibfnamefont{K.~K.}
  \bibnamefont{Perkins}}, \emph{\bibinfo{title}{Transforming the junior level:
  Outcomes from instruction and research in e\&m}}, \bibinfo{journal}{Phys.
  Rev. ST Phys. Educ. Res.} \textbf{\bibinfo{volume}{8}},
  \bibinfo{pages}{020107} (\bibinfo{year}{2012}).

\bibitem[{\citenamefont{Wilcox and Pollock}(2014)}]{wilcox2014cmr}
\bibinfo{author}{\bibfnamefont{B.~R.} \bibnamefont{Wilcox}} \bibnamefont{and}
  \bibinfo{author}{\bibfnamefont{S.~J.} \bibnamefont{Pollock}},
  \emph{\bibinfo{title}{Coupled multiple-response versus free-response
  conceptual assessment: An example from upper-division physics}},
  \bibinfo{journal}{Phys. Rev. ST Phys. Educ. Res}
  \textbf{\bibinfo{volume}{10}}, \bibinfo{pages}{020124}
  (\bibinfo{year}{2014}),
  \urlprefix\url{http://link.aps.org/doi/10.1103/PhysRevSTPER.10.020124}.

\bibitem[{\citenamefont{{See Supplemental Material at http://link.aps.org/
  supplemental/10.1103/PhysRevSTPER.10.020124}}(for full multiple-response CUE
  post-test)}]{q8cmr}
\bibinfo{author}{\bibnamefont{{See Supplemental Material at
  http://link.aps.org/ supplemental/10.1103/PhysRevSTPER.10.020124}}}
  (\bibinfo{year}{for full multiple-response CUE post-test}).

\bibitem[{\citenamefont{Catrambone}(2011)}]{catrambone2011taps}
\bibinfo{author}{\bibfnamefont{R.}~\bibnamefont{Catrambone}}, in
  \emph{\bibinfo{booktitle}{2011 Learning and Technology Symposium}}
  (\bibinfo{address}{Columbus, Georgia}, \bibinfo{year}{2011}).

\end{thebibliography}
\bibliographystyle{apsper}   

\end{document}